\documentstyle [12pt] {article}

\begin{document}
\newcommand{\ra}{\rightarrow}
\newcommand{\cp}{\chi^\prime_{cJ}}
\newcommand{\pp}{\psi^\prime}
\newcommand{\T}{P_T}
\newcommand{\bp}{p\bar{p}}
\newcommand{\J}{J/\psi}
\newcommand{\bD}{D \bar{D}}
\newcommand{\CJ}{\chi_{cJ}}
\vspace{0.5in}
\oddsidemargin -.070in
\newcount\sectionnumber
\sectionnumber=0
\def\be{\begin{equation}}
\def\ee{\end{equation}}
\vspace {.5in}
\begin{center}
{\Large\bf Search for the $\chi'_c$ Charmonium States as
Solution to the CDF $\psi'$ Puzzle. \\}
\vspace{.5in}
{\bf S.F. Tuan \\}
\vspace{.1in}
 {\it
Physics Department, University of Hawaii at Manoa, 2505 Correa
Road, Honolulu, HI 96822, USA.}\\

[Submitted to Pramana (4/10/95)] \\

\end{center}

\vskip .1in
\begin{abstract}

The efforts of Roy-Sridhar-Close-Cho-Wise-Trivedi to resolve the CDF
$\psi'$ anomaly with cascades from above-threshold
$\chi^\prime_c$
states require well defined signatures [a small total width
and a large branching fraction for $\cp \ra \gamma + \pp]$
for the
solution to be viable.  Here we estimate the production of
such states from $BR (B \ra \cp + X)
BR (\cp \ra \gamma \pp)$ and $\gamma \gamma$ production of
$\chi^\prime_{c2}$ at CLEO II, and comment on the
feasibility of testing the hypothesis in terms of current
experimental capabilities.
\end{abstract}
\vspace{.50in}
\vskip .25in
\section{Introduction}

The CDF measurement \cite{1} of $\J$ and $\pp$ production at large
transverse momentum $(\T)$ in 1.8 TeV $\bp$ collisions at the
Tevatron has produced one of the most intriguing
experimental results in recent times.
At low transverse momentum the production of $\psi$'s in $\bp$
in collisions is
expected to proceed via the production of $\chi$ states
followed by their decay \cite{2}, i.e. fusion process $g + g \ra \chi
\ra \psi + X$.  The analogous
process at larger transverse momentum is $gg \ra g \psi$.  This process
generates a cross section that falls off at large $\T$
much faster than, say, the jet rate.  This observation led
to an expectation that $\psi$'s produced at large $\T$  came almost
exclusively from b-quark decay.  By detecting whether or not
the $\psi$'s come from the primary event vertex, CDF has tested
this expectation.  The fraction of $\psi$'s produced directly is
almost independent of $\T$  and the rate of direct $\psi$ production
at large $\T$ is (substantially) larger that had been expected.
This expectation is thus now known to be {\underline{false}}.

The dominant production mechanism of $J/\psi$'s at large tranverse
momentum is now believed to be the fragmentation of light
(and charm) quark and gluon jets into $\chi_c$ mesons that
subsequently decay
to $\J$ \cite{2} via the radiative mode as stressed by Cho, Wise,
and Trivedi \cite{3} where the gluon fragmentation is found to be
particularly important.  When this $\J$ source is included, the
theoretical prediction for $d \sigma ( \bp \ra \J +
X)/dP_T$ at $\sqrt{s} = 1.8$ TeV agrees within a
factor of 2 with recent CDF data \cite{1}.

While the rate for $J/\psi$ production is in agreement with
expectations, given the inherent theoretical uncertainties,
the rate for $\psi^\prime$ production at CDF is at least a factor
of 20 above theoretical expectations \cite{4}.  The calculation
does not include the possibility of $\psi'$ production from
the decay of $\chi^\prime_c$, 2P states.  These 2P charmonium
states are above
$\bD$ threshold, however a branching ratio of a few \% to $\psi'$
could be sufficient to explain the deficit.  D.P. Roy and K.
Sridhar as well as others investigated this possibility
quantitatively \cite{3}.

The basic premise is to recall that $\psi'$ is the heaviest
$c \bar{c}$ bound state which lies below the $\bD$ threshold.
Therefore,
n=1 $\CJ$ states cannot radiatively decay to $\pp$  but their n=2
counterparts can.  None of these $\cp$ (or $\chi_{cJ}$ (2P))
states which lie above
the $\bD$ threshold have been observed.  Estimates of their
masses yield $M (\chi_{c0} (2P)) = 3920 \ \rm{MeV}, M
(\chi_{c1}(2P) =3950$ MeV, and $ M (\chi_{c2} (2P)) =3980$ MeV \cite{5}.
            These mass values taken literally
would kinematically allow the S-wave transitions
$\chi_{c0}(2P) \ra \bD$ and $\chi_{c1} (2P) \ra \bD$ to
occur.  We therefore expect that the J=0 and perhaps
the J=1 excited $\CJ$ (2P) states will be broad and have negligible
branching fractions to lower $c \bar{c}$ bound states.  However,
angular momentum and parity considerations require the
analogous decays $\chi_{c2}$ (2P) $\ra D \bar{D}$ and
$\chi_{c2} (2P) \ra D^*
\bar{D}$ for the J=2
state to proceed via L=2 partial waves.
Although we cannot readily compute by how
much these D-wave decays will be suppressed, it is possible
that the branching fractions for $\chi_{c2}$ (2P) states to charmonium
states below $D\bar{D}$ threshold could be significant.  F.
Close \cite{3} suggests that the $\chi_{c1}$ (2P) may also
have suppressed hadronic widths due to quantum numbers or
nodes in form factors manifested in decays near threshold,
e.g. since $^3P_1 \ra \bar{D}D^*$ is near threshold the $S$ and $D$ waves
present are affected by radial wave function nodes which can
conspire to reduce the width.  Hence in section 2 below  we
will consider the search for both $\chi_{c1} (2P)$ and
$\chi_{c2} (2P)$.  Although the $^1D_2$ and $^3D_2$
charmonium states may be present in the CDF data and
detectable, they are unlikely to explain the $\psi^\prime$ (3685)
enhancement.  For instance \cite{3} $^1D_2$ production is
suppressed, and it is expected to have a very small
branching ratio to $\psi^\prime \gamma$.

On a  quantitative basis Page \cite{6} has calculated the
total
widths of radial $\cp$ states and found that they could be as
small as 1-5 MeV.  We need an estimate of the branching
ratio $BR (\cp \ra \gamma + \psi^\prime)$.
A very rough estimate \cite{7} can be obtained using
the known experimentally
measured branching ratio $BR (\chi_{b2} (2P) \ra \Upsilon
(2S) + \gamma) \approx 16\%$ \cite{8} as follows:
\be
BR (\chi_{c2} (2P) \ra \gamma \pp ) =
\left ( \frac{Q_c}{Q_b} \right )^2
\left ( \frac{k_c}{k_b} \right )^3 BR (\chi_{b2} (2P) \ra
\gamma \Upsilon(2S))
\frac{\Gamma_{tot} (\chi_{b2} (2P))}
{\Gamma_{tot} (\chi_{c2} (2P))}
\ee
Here $(Q_c/Q_b)^2 = \frac{4/3}{1/3} = 4,$  and $(k_c/k_b)^3$ is the
modification due to $E_1$ phase space; there
will be some changes due to the actual size of $b \bar{b}$
relative to that of $ c \bar{c}$
but this will be a factor of 2 or 3 and the estimate (1) may
not be accurate to better than a factor of 5 to 10 anyway.
 For instance, taking $M(\chi_{c2}$ (2P)) = 3980 MeV and
$\Gamma_{tot} (\chi_{b2} (2P)) \approx \Gamma_{tot}
(\chi_{c2}$ (2P)) gives $BR (\chi_{c2} (2P)
\ra \gamma + \psi^\prime) \approx 100\%!$  Thus the value $BR
(\chi_{c2} (2P) \ra \gamma + \pp ) \approx 10\%$ suggested by
Roy and Sridhar \cite{3} is by no means
unreasonable.  Another back of the envelope ansatz \cite{7} is to
take the $1P \ra 1S$ charmonium data and
{\underline{assume}} the 2P $\rightarrow
2S$ overlaps will have the same order of magnitude, then $B
(\chi_{c2} (2P) \ra \gamma + \psi^\prime)
\approx B
(\chi_{c2} (1P) \ra \gamma + \J$) = 13.5\%, with a measured
$\Gamma_{c2} (1P)$ full width = 2 MeV \cite{8}.
Hence if one of the $\chi_{cJ}$ (2P) states is
calculated to have a total width in the range 1 MeV
to 5 MeV \cite{6}, a branching ratio $B (\chi_{cJ} (2P) \ra \gamma + \pp )
>$ 5\% can be expected (a value in the range
5-10\% would be needed to explain the CDF
$\psi'$ anomaly \cite{1,3}).  To summarize, the result is that in order of
magnitude one expects the radiative transition to be
\cal{O}(100
KeV) and hence the BR is \cal{O}(1-10\%) if the hadronic width is
\cal{O}(10-1MeV).  It would be surprising if the branching ratio is
less than 1\% or much geater than 10\%.  To give an adequate
spread for illustration, we shall take in section 2,

\be
BR (\chi_{cJ} (2P) \ra \gamma + \psi (2S)) = 1, 5, 10\%.
\ee

\section{Search Method for $\chi'_{cJ}$ States}

The optimal method for accumulation of $\cp$ events at CLEO
II  is to take advantage of the inclusive decays of B mesons
to Charmonium.  Hence we seek to estimate
\be
B_J \equiv BR [B \ra \cp + X] \times B [ \cp \ra \gamma \pp
].
\ee
The branching ratio $BR [B \ra \cp  + X ]$
is given by Bodwin et al. \cite{9} as
\be
R (\CJ) \times 10.7\% \times \mid R^\prime_{\chi^\prime_{c}}
(0)/ R^\prime_{\chi_{c}} (0) \mid^2.
\ee
where $R(\chi_{cJ}) = \Gamma (b \ra \chi_{cJ} + X) /
\Gamma (b \ra e^- \bar{\nu}_e + X),$ 10.7\%
is the observed semileptonic branching
ratio for the B-meson, and multiplicative last term
$\mid R^\prime_{\chi^\prime_c} (0)/R^\prime_{\chi_c} (0)
\mid^2$ is
Braaten's correction factor \cite{10} for estimating $BR(B
\rightarrow \chi'_c + X)$ from $BR (B \ra \chi_c + X)$.
The derivatives of the
wave functions at the origin can be obtained from potential
models.  This procedure is certainly correct for the
color-singlet matrix element (the $c\bar{c}$ contribution) and it
may also be correct for the color octet matrix element (the
$c\bar{c}g$ contribution) if the latter is dominated by the radiation
of soft gluons from the $c\bar{c}$ state.  Whether or not this is
the case remains to be seen \cite{10}.

The expression $R (\chi_{cJ})$ in (4) is given in terms of the
nonperturbative parameters $H_1$ and $H^\prime_8$ (proportional
to the
probabilities for a $c \bar{c}$ pair in a color-singlet P-wave and a
color-octet S-wave state, respectively, to fragment into a
color singlet P-wave bound state) and takes the following
form \cite{9}
\begin{eqnarray}
\begin{array}{l}
R(\chi_{c1}) \cong 12.4 \ (2C_+ - C_-)^2 \ H_1/M_b + 9.3 (C_+ +
C_-)^2 \ H^\prime_8 (M_b)/M_b
\\
R (\chi_{c2}) \cong 15.3 \ (C_+ + C_-)^2 \ H^\prime_8 (M_b) /M_b.
\end{array}
\end{eqnarray}

Here $C_+$ and $C_-$ are Wilson coefficients that arise from
evolving the effective 4-quark interactions mediated by the
W boson from the scale $M_W$ down to the scale $M_b$.
Numerically $C_+ (M_b) \cong 0.87$, $C_-(M_b) \cong 1.34, H_1
\approx 15$ MeV, $H^\prime_8 (M_b) \cong 2.5$ MeV, and $M_b
= 5.3$ GeV

For the value $\mid R^\prime_{\chi_{c}^\prime}
(0)/R^\prime_{\chi_{c}} (0)\mid^2$, we use a recent quark potential
model \cite{11} which takes into account that the value of
$R^\prime (0)$ for
P-state charmonium is sensitive to the short distance
behavior of the potential, so that it is better to use
values obtained from potentials whose short distance
behavior is more reliable.  The potential \cite{11} (an improved
version of the Buchm$\ddot{\rm{u}}$ller-Grunberg-Tye potential) approaches
the 2-loop QCD result at short distance, leads to
energy spectra and leptonic widths in very good agreement
with experiment.  The values for $R^\prime(0)$ for the P-wave
charmonium states from this potential model Program \cite{12}
are:
\begin{eqnarray}
\begin{array}{cc}
State  &  \quad R^\prime(0) \ \ \mbox{in} \ \mbox{GeV}^{5/2}  \\
\chi_c &   0.20   \\
\chi^\prime_c      &    0.23
\end{array}
\end{eqnarray}

For comparison, the value of $R^\prime(0)$ for
$\chi_c$ \cite{10}
obtained directly from the
measured widths of $\chi_{c1}$ and $\chi_{c2}$ was about
0.15 GeV$^{5/2}$, hence one should ascribe an error of
not less than 30\% to any of
these values.  It is nevertheless reassuring that a recent
compilation \cite{13} of first nonvanishing derivative
at zero $c \bar{c}$
separation for radial Schr$\ddot{o}$dinger wave function of earlier
potential models, give for $\mid
R^\prime_{\chi_{c}^\prime}(0)/R^\prime_{\chi_{c}}(0)\mid^2$
values 1.36 (Buchm$\ddot{\rm{u}}$ller-Tye), 1.05 (Power-law), 0.97
(Logarithmic), and 1.42 (Cornell), quite compatible with
\be
\mid \frac{R^\prime_{\chi_{c}^\prime} (0)}{R^\prime_{\chi_{c}}
(0)} \mid ^2 \  = \ 1.32
\ee
of the recent potential model \cite{11}.

Assembling the pieces together from (3), (4), (5), and (7),
we have for (3) and J=1.
\be
B_1 = 2.89 \times 10^{-5}, 1.45 \times 10^{-4}, 2.89 \times
10^{-4}
\ee
for the respective  values of $BR [\chi^\prime_{c1} \ra
\gamma \psi^\prime]$ given in (2).
The corresponding values for J=2 are
\be
B_2 = 3.77 \times 10^{-5}, 1.89 \times 10^{-4}, 3.77 \times
10^{-4}
\ee
The world average (WA) and CLEO-II measurement \cite{14} for
$BR(B \ra \chi_c + \ X)$ are
\begin{eqnarray}
BR (B \ra \chi_{c1} + X) &=& 0.42 \pm 0.07\% \ \ (WA) \\
\nonumber
BR (B \ra \chi_{c2} + X) &=& 0.25 \pm 0.10\% \ \ (CLEO)
\end{eqnarray}

For central values of (10), multiplying the above
experimental numbers by correction factor (7), we have
\begin{eqnarray}
BR (B \ra \chi^\prime_{c1} + X) &=& 5.54 \times 10^{-3} \\
\nonumber
BR (B \ra \chi^\prime_{c2} + X) &=& 3.30 \times 10^{-3}
\end{eqnarray}

This compares with the values obtained by scaling the
predictions of Bodwin et al. \cite{9} for $BR (B \ra \CJ +
X)$ using the correction factor (7) of
\begin{eqnarray}
BR (B \ra \chi^\prime_{c1} + X) &=& 2.89 \times 10^{-3} \\
\nonumber
BR (B \ra \chi^\prime_{c2} + X) &=& 3.77 \times 10^{-3} .
\end{eqnarray}
The agreement seems good for the $\chi^\prime_{c2}$ case.

The final step is to estimate the number of observed events,
$N^{obs}_J$.
First we note that at CLEO-II, the number of
produced B-mesons is given by
\be
N_B^{\mbox{produced}} = (\int {\cal L} dt) \times \sigma
(e^+e^- \ra \Upsilon (4S) \ra B \bar{B}) \times 2
\ee
where on a good year the integrated luminosity $\int {\cal L}
dt$ is $2fb^{-1}$ of
data on tape, the $\sigma (e^+e^- \ra \Upsilon (4S) \ra B
\bar{B})$ cross section is about 1.07 nb, and the
factor of 2 takes into account productions of pairs of
B mesons in $\Upsilon (4S)$ decays.  The number of observed events
$N^{obs}_J$
is then given by
\be
N^{obs}_{J} = N_B^{\mbox{produced}} \times B_J \times B
(\psi^\prime \ra \J \pi^+ \pi^-) \times (\sum_{\ell=e, \mu}
B ( \J \ra \ell^+ \ell^-)) \times \epsilon
\ee
where $B_J$ is defined in (3), and $\epsilon$
is the efficiency for detecting $\psi^\prime \ra \J
\pi^+\pi^-$ in the dilepton mode
(about 20\% in the CLEO-II detector \cite{14}).  Hence using theory (8),
(9), or (12), we have
\begin{eqnarray}
N_1^{obs} & = & 0.95, \ 4.77, \ 9.51 \ \mbox{events}  \\
\nonumber
N_2^{obs} & = & 1.24, \ 6.22, \ 12.40 \ \mbox{events}
\end{eqnarray}
for the respective values of $BR(\cp \ra \gamma \pp)$
given in (2).  If
we take advantage of the experimentally known branching
ratios (10) in deducing (11), we have
\begin{eqnarray}
N_1^{obs} & =  1.82, \ 9.11, \ 18.23 \ \mbox{events}  \\
\nonumber
N_2^{obs} & =  1.09, \ 5.43, \ 10.86  \ \mbox{events}
\end{eqnarray}
for the respective values of $BR (\cp \ra \gamma  \pp)$.
The theoretical mass
estimates \cite{5}  $M(\chi_{c1} (2P)) =3950$ MeV
and $M(\chi_{c2} (2P)) = 3980$ MeV are also useful.
The approximate locations of these resonances are needed to
conduct the experimental search and reduce background.

At CLEO II, the $\chi^\prime_{c2}$ state can also be
searched for via the two photon production of this J=2
state.  For instance, the number of events
$N_{\chi^\prime_{c2}}$ from $e^+e^- \ra e^+e^- \gamma \gamma
\ra e^+e^-\chi^\prime_{c2}$ is estimated to be
\begin{eqnarray}
\begin{array}{lclrcl}
N_{\chi^\prime_{c2}} & = & N_{\chi_{c2}} & \times
\frac{\sigma (\gamma \gamma \ra \chi^\prime_{c2})}
{\sigma (\gamma \gamma \ra \chi_{c2})}
& \times    &
\frac{BR (\chi^\prime_{c2} \ra \psi^\prime \gamma)}
{BR (\chi_{c2} \ra J/\psi\gamma)} \times
\\
 \quad  & \times & \sum_{\ell=e,\mu}
   & \frac{\epsilon (\ell^+ \ell^- \pi^+ \pi^- \gamma)}
         {\epsilon (\ell^+ \ell^-\gamma) } &
\times   &
\frac{BR( \psi^\prime \ra J/\psi \pi^+ \pi^-) \times BR
(J/\psi \ra \ell^+ \ell^-)}
{BR(J/\psi \ra \ell^+ \ell^-)}
\end{array}
\end{eqnarray}

In a recent paper on measurement of two-photon production of
the $\chi_{c2}$, J. Dominick et al. [15], using 1.5
fb$^{-1}$ of data taken with beam energies near the
$\Upsilon(4S)$, 25.4 $\pm 6.9 \ N_{\chi_{c2}}$ events were
obtained, with efficiency $\epsilon(\ell^+\ell^- \gamma)$  = 0.187
$\pm$ 0.003.  Taking into account that data accumulation is
now 2 fb$^{-1}$ on $\Upsilon(4S)$ and 1 fb$^{-1}$ in the
continuum, the number $N_{\chi_{c2}}$ can be doubled and we
assume that $\epsilon (\ell^+ \ell^- \pi^+ \pi^- \gamma)
\cong \epsilon (\ell^+ \ell^- \gamma)/2$.  For
$BR(\chi^\prime_{c2} \ra \psi' \gamma) = 10\%$ and taking
central values for experimental numbers to illustrate, we
have
\be
N_{\chi^\prime_{c2}} \cong 6 \times
\frac{\sigma (\gamma \gamma \ra \chi^\prime_{c2})}
{\sigma (\gamma \gamma \ra \chi_{c2})}  \ \mbox{events}.
\ee

The cross section ratio is equal to the $BR(\chi^\prime_{c2}
\ra \gamma \gamma)/BR(\chi_{c2} \ra \gamma \gamma)$ and is
not yet known since the total width $\Gamma
(\chi^\prime_{c2})$ has not yet been measured.

\section{Conclusions}

We have presented above in section 2 event rates for
observing the $\chi^\prime_{cJ} (J=1,2)$ in B decays at CLEO-II.
Though the event rates $N_J^{obs}$  given by (15) and
(16)
are  not large even with a year's accumulation of
$B\bar{B}$,
they can be steadily increased by extending the
$B\bar{B}$ accumulation
over a period of several years.  Furthermore we have been
surprised by how large the branching ratios $BR(\cp \ra
\gamma \pp )$ can be in
section 1, given the known and sizable \cite{8} $BR (\chi_{b1}
\ra \gamma + \Upsilon (2S)) \approx (21 \pm 4\%)$
and $BR (\chi_{b2} \ra \gamma + \Upsilon (2S)) \cong (16.2
\pm 2.4) \%$.  Hence
optimistically we can expect $N_J^{obs}$ to be in the
range of 10-20
events for an integrated luminosity of $2 fb^{-1}$, as needed to
explain the CDF $\psi^\prime$ anomaly \cite{1}.
Estimates for $2 \gamma$ production of $\chi^\prime_{c2}$ are
given in (17) and (18).  At CDF the invariant mass spectrum
of $\gamma \psi'$ combination can be studied. This
possibility should
be explored in parallel with the CLEO effort.
I wish to thank  D.P. Roy for encouragement, my colleagues
X. Tata and T. Browder for their comments and reading of the
manuscript, and E. Braaten, F. E. Close, Y. P. Kuang, and P.
Page for very helpful communications.  This work was
supported in part by the US Department of Energy under Grant
DE-FG-03-94ER40833.

\end{document}